\makeatletter
\let\@twosidetrue\@twosidefalse
\let\@mparswitchtrue\@mparswitchfalse
\makeatother

\documentclass[12pt]{llncs}
\usepackage[a4paper, left=3.1cm, right=3.1cm, top=2.9cm, bottom=2.9cm]{geometry}

\usepackage{graphicx}
\usepackage{amsmath, amssymb}

\usepackage{natbib}
\usepackage[dvipsnames, table]{xcolor}
\definecolor{oneblue}{rgb}{0.0, 0.0, 0.85}
\definecolor{denimblue}{rgb}{0.08, 0.38, 0.74}

\usepackage[colorlinks,
           urlcolor=oneblue,
           linkcolor=denimblue,
           citecolor=NavyBlue,
           bookmarksopen=false,
           pdfpagemode=UseNone,
           pagebackref]{hyperref}

\usepackage{mathtools}
\usepackage[nottoc,notlof,notlot]{tocbibind}

\newcommand{\up}[1]{$^{\mathrm{\textsf{#1}}}$} 

\newcommand{\ue}{\mathrm{e}}
\newcommand{\ui}{\mathrm{i}}
\renewcommand{\geq}{\geqslant}

\begin{document}


\title{Tsunami hazard assessment in the Makran subduction zone}

\author{Amin \textsc{Rashidi}\inst{\,1}\thanks{Corresponding author.\  E-mail address: amin.rashidi@ut.ac.ir (A.~\textsc{Rashidi}).}, Zaher Hossein \textsc{Shomali}\inst{\,1,2}, Denys \textsc{Dutykh}\inst{\,3,4} and Nasser \textsc{Keshavarz Farajkhah}\inst{\,5}}
\authorrunning{Amin Rashidi \emph{et al}.} 
\institute{Institute\ of Geophysics, University\ of Tehran, Tehran, Iran\\
\and
Department of Earth Sciences, Uppsala University, Uppsala, Sweden
\and
Univ. Grenoble Alpes, Univ. Savoie Mont Blanc, CNRS, LAMA, 73000 Chamb\'ery, France
\and
LAMA UMR 5127 CNRS, Universit\'e Savoie Mont Blanc, Campus Scientifique, 73376 Le Bourget-du-Lac, France \\
\and
Research Institute of Petroleum Industry, Tehran, Iran}

\maketitle              


\begin{abstract}
The lack of offshore seismic data caused uncertainties associated to understating the behavior of future tsunamigenic earthquakes in the \textsc{Makran} Subduction Zone (MSZ). Future tsunamigenic events in the MSZ may trigger significant near-field tsunamis. Run-up heights in the near-field are controlled by the heterogeneity of slip over the rupture area. Considering a non-planar geometry for the \textsc{Makran} subduction zone, a range of random $k^{\,-2}$ slip models were generated to hypothesize rupturing on the fault zone. We model tsunamis numerically and assess probabilistic tsunami hazard in the near-field for all synthetic scenarios. The main affected areas by tsunami waves are the area between \textsc{Jask} and \textsc{Ormara} along the shorelines of \textsc{Iran} and \textsc{Pakistan} and the area between \textsc{Muscat} and \textsc{Sur} along the \textsc{Oman} coastline. The maximum peak-run-up along the shores of \textsc{Iran} and \textsc{Pakistan} is about 16 m and about $12$ $\mathsf{m}$ for the \textsc{Oman} shoreline. The slip distributions control the run-up along the \textsc{Makran} coastlines. The dependency of run-up to the heterogeneity of slip is higher in the most impacted areas. Those areas are more vulnerable to tsunami hazard than other areas. 

\end{abstract}

\section{Introduction}

The \textsc{Makran} Subduction Zone (MSZ) is a convergent plate boundary between the \textsc{Arabian} plate and the overlying \textsc{Eurasian} plate with an average convergent rate of $\sim 4$ $\mathsf{cm/yr}$. It extends about $900$ km from southeastern \textsc{Iran} to southern \textsc{Pakistan}. The 1945 \textsc{Makran} tsunami and the possibility of generating future tsunamigenic events by the \textsc{Makran} subduction zone have motivated researches recently to study tsunami hazard in the northwestern \textsc{Indian Ocean} \citep[\emph{e.g.}][]{Okal2008, Heidarzadeh2008, Rajendran2008, Heidarzadeh2011, Rehman2013, El-Hussain2016, nhess-16-1339-2016}. 

The occurrence of the $1945$ \textsc{Makran} and the evidences of earlier historic earthquakes (\emph{e.g.} $1008$ and $1483$) have proved earthquake and tsunami potential of the \textsc{Makran} subduction zone \citep{JGRB:JGRB8463, Heidarzadeh2008, GRL:GRL50374}. It is seismically active, being capable of generating large earthquakes. However, it is segmented into the western and eastern parts from the seismic potential point of view. The active eastern segment generated the $1945$ tsunamigenic earthquake ($M_{\,w}$ $8.1$) which caused a regional tsunami with maximum wave heights of $10$ $\mathsf{m}$ and about $4\,000$ human losses \citep{HEIDARZADEH2008774, Rajendran2013}. The potential of the western segment for generating tsunamigenic earthquakes is puzzling due to its apparent aseismictiy and the lack of historical data. There has been some confusion in the studies on which behavior can explain the current status of the western \textsc{Makran}. The presence of sediment layers may be the cause of unusual seismic behavior of the western \textsc{Makran}. Until ​the ​end of 2004, previous investigations have suggested that the \textsc{Sumatra} subduction zone, at least the northern part, was considered to be ​sleeping from which tsunamigenic events would not be expected.​ The lack of historic data on major seismic events during the last $200$ \textsf{years}​ was supporting this misconception. The gigantic 2004 \textsc{Sumatra Andaman} earthquake strongly disproved this ​belief.​ After more thorough ​investigations, the traces of major tsunamis have been found in sediment deposits on the horizon of $1\,000$ \textsf{years}. The same story ​may​ happen to the MSZ, especially for the western segment.​ The duration of our seismic records and observations is not enough to predict reliably earthquakes on long term. Another example is the recent \textsc{T\={o}hoku} 2011 tsunamigenic event. The \textsc{Fukushima} nuclear plant was protected against tsunamis, but the chosen tsunami design height​ was twice smaller than the real wave which hit the plant. One can assume that the western segment is currently locked and capable of generating future plate boundary earthquakes \citep{doi:10.1144/0016-76492008-119, Rajendran2013}. This assumption leads to ​serious reconsider​ation​ of the entire \textsc{Makran} rupturing​ scenario​. \textsc{Makran} subduction zone is characterized as a shallow subduction zone which can increase coupling and provide a suitable condition to release large earthquakes \citep{10.1007/978-3-540-87974-9_7, JGRB:JGRB17250}.

The $2004$ \textsc{Indian Ocean} and $2011$ \textsc{T\={o}hoku} tsunamis have demonstrated that the scale and impacts of tsunamis can be significantly larger than expected \citep{Suppasri2013, Satake2014_2} showing the need of increasing attention to the hazard potential of subduction zones for worst case scenarios. 

The effect of far-field tsunamis can be suitably evaluated knowing the seismic moment \textit{M}$_0$ of the submarine earthquake \citep{Dutykh2012a}. However, the tsunami hazard of tsunamigenic earthquakes in the near field depends on several factors which have increased the researchers interest recently \citep[\emph{e.g.}][]{Dutykh2011c, Lay2011, GRL:GRL28100, GRL:GRL28538, Dutykh2012a}. Although, the tsunami propagation \emph{per se} is well modeled and understood, it can be still problematic due to uncertainties in the rupture details \citep{Greenslade2008}. As a lesson from the $2004$ \textsc{Indian Ocean} and $2011$ \textsc{T\={o}hoku} tsunamis, the tsunami generation process and its relationship with seismogenic subduction zones need to be investigated in more detail \citep{Fujii2011, Ide1426}. The tsunami generation simulation is the controlling step in modeling the life stages of tsunamis and their effects on coastal areas. In the case of tectonic-generated tsunamis, rupture geometry, focal mechanism, seismic moment and coseismic slip distribution describe the tsunami generation process. They are required to compute the $3$D seafloor deformation caused by rupturing on an underwater fault. Tsunami numerical modeling may fail to predict the true effects of tsunamis by assuming simple uniform slip distributions. Heterogeneity of slip over the fault controls local tsunami wave amplitude variations \citep{Geist1999}, which is necessary to model near-field tsunamis accurately and assess their hazard.

Tsunami hazard can be assessed in three different approaches \citep{JGRC:JGRC11257}: (1) Probabilistic Tsunami Hazard Assessment (PTHA), (2) worst-case scenario approach (deterministic), and (3) sensitivity analysis. The last one \citep[see \emph{e.g.}][]{article_Tang, BARKAN2009109} studies the sensitivity of flooding or run-up to the characteristics of tsunami sources which does not consider the probability of each scenario \citep{JGRC:JGRC11257, Dutykh2011c}.

Deterministic approach \citep[\emph{e.g.}][]{JGRB:JGRB15310, article_Pr} considers particular scenarios (usually the worst case scenario) to calculate their effect on specific areas. Probabilistic Tsunami Hazard Assessment (PTHA) provides a useful tool to evaluate tsunami risk by considering the probabilities of tsunamis. It determines the likelihood of tsunami severity usually in terms of tsunami run-up or flooding for a range of possible sources. PTHA computes the probability of tsunami run-up or inundation at a given location that exceeds a certain level in a specified time period. It was developed by \cite{Lin1982}, \cite{Rikitake1988:} and \cite{Downes2001} by modifying the probabilistic seismic hazard assessment (PSHA). Increased attention has been paid recently to PTHA, especially after the $2004$ \textsc{Indian Ocean} tsunami \citep[\emph{e.g.}][]{Geist2006, JGRC:JGRC11257, nhess-12-151-2012, JGRB:JGRB16832, nhess-14-3105-2014}. PTHA in \textsc{Makran} has been also a subject of interest to some researchers \citep[\emph{e.g.}][]{Heidarzadeh2011, El-Hussain2016, nhess-16-1339-2016}.

In our previous study \citep{Rashidi2018}, the evolution of kinetic, potential and total energies in the near-field was studied for a multi-segment source model of the \textsc{Makran} subduction zone. An empirical relationship between the moment magnitude and tsunami wave energy in the \textsc{Makran} subduction zone was obtained. In the current study, we model hypothetical earthquakes generated by the entire \textsc{Makran} subduction zone and perform PTHA. Heterogeneous slip distribution patterns are considered to represent the complexity of earthquakes. Spatial non-uniform slip models are a better approximation of the rupture complexity to model seismic sources \citep{Ruiz2015}, however, it is ignored in many tsunami studies assuming simple source models \citep{Geist2002}. The current study is focused on the $k^{-2}$ rupture model, introduced by \cite{Bernard1994} and developed by \cite{Gallovič2004}. The distributions of run-up along the shores of \textsc{Iran}, \textsc{Pakistan} and \textsc{Oman} are used in our analysis and assessment.
%
%
\section{Methodology}

\subsection{Earthquake source models}

Heterogeneity of slip in plate boundary events greatly impacts the vertical seabed displacement and thus the tsunami generation \citep{Geist2002}. The lack of offshore seismic data has limited our understanding of the present-day behavior of the \textsc{Makran} subduction zone and insight into the complexity of slip distribution of possible future events. Alternatively, synthetic slip models can be useful tools to exhibit the complexity of tsunamigenic sources and to understand their effects in the \textsc{Makran} region. Using $k^{\,-2}$ stochastic source models, we generate digital static tsunamigenic ruptures to model near-field tsunamis. \cite{Bernard1994} proposed the so-called \emph{kinematic self-similar rupture model} in which spectral amplitudes of a random slip distribution decay as power of $2$ at high wave numbers $k$ beyond the corner wavenumber $k_{\,c}\ =\ 1/L_{\,c}\,$, where $L_{\,c}$ is the characteristic dimension of the fault (usually the length) \citep{Gallovič2004}. The slip does not depend on $k$ below $k_c$. In the case of a rectangular fault with a lenght of $L$ and a width of $W$, the $2$D slip distribution $D\,(k_{\,x},\,k_{\,z})$ can be written as its spatial \textsc{Fourier} spectrum \citep{Gallovič2004}
%
\begin{equation} \label{eq:Slp}
D\,(k_{\,x},\,k_{\,z})\ =\ \frac{\mathrm{\Delta}\bar{u}\;L\, W}{\sqrt{1\ +\ \biggl(\Bigl(\dfrac{k_{\,x}\,L}{K}\Bigr)^{\,2}\ +\ \Bigl(\dfrac{k_{\,z}\,W}{K}\Bigr)^{\,2}\biggr)}} \; \ue^{\,\ui\,\Phi\,(k_{\,x},\,k_{\,z})}\,,
\end{equation}
where $\mathrm{\Delta}\bar{u}$ is the mean slip and $\Phi\,(k_{\,x},\,k_{\,z})$ is the phase spectrum, $k_{\,x}$ and $k_{\,z}$ are the wavenumber components along the $x$ and $z$ directions and $K$ is a dimensionless constant which determines the smoothness of slip distribution \citep{Gallovič2004}. Based on \cite{doi:10.1785/gssrl.70.1.59}, $K\ <\ 1$ is suggested, which corresponds to smoother slip distributions. 

\begin{figure}
\centering
\includegraphics[scale=3]{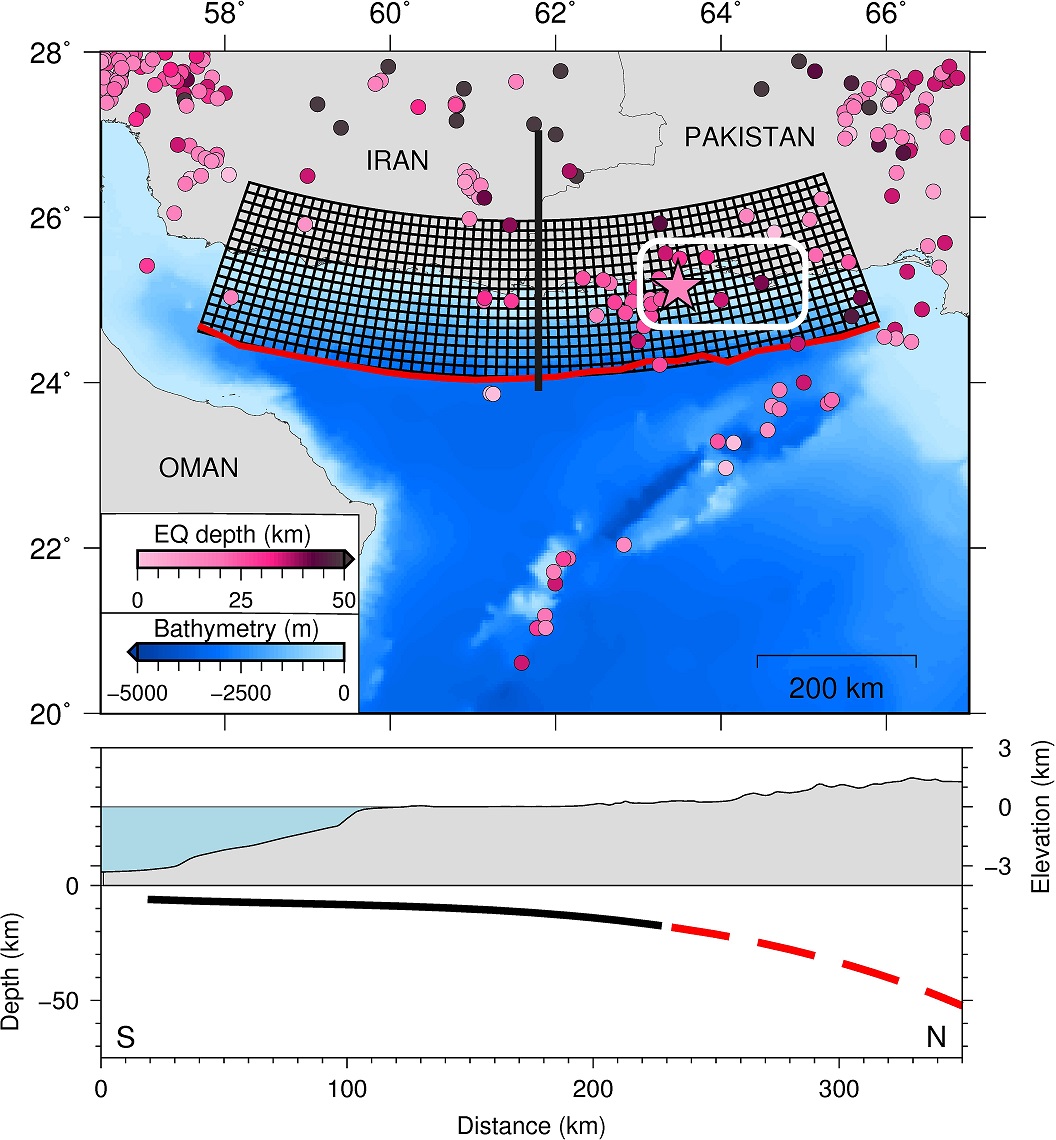}
\caption{\small\em Upper panel: Seismicty of the \textsc{Makran} subduction zone (for \textit{M}4.0+ and between 1926 to 2016 from the ISC catalog). The star shows the epicenter of $1945$ $M_{\,w}$ $8.1$ \textsc{Makran} earthquake. The white outline represents the estimate of the approximate rupture area of the $1945$ event \citep{JGRB:JGRB8463}. The color of circles indicates the depth of events. The red line is the deformation front by \cite{GRL:GRL50374}. The thick black line orthogonal to the deformation front denotes the cross-section. The non-planar mesh is the rupture geometry. Lower panel: Cross-section of the fault geometry (black line) and the palte interface (dashed red line) from \cite{GRL:GRL50374}.}
\label{fig:mak}
\end{figure}

\begin{figure}
\centering
\includegraphics[scale=2.3]{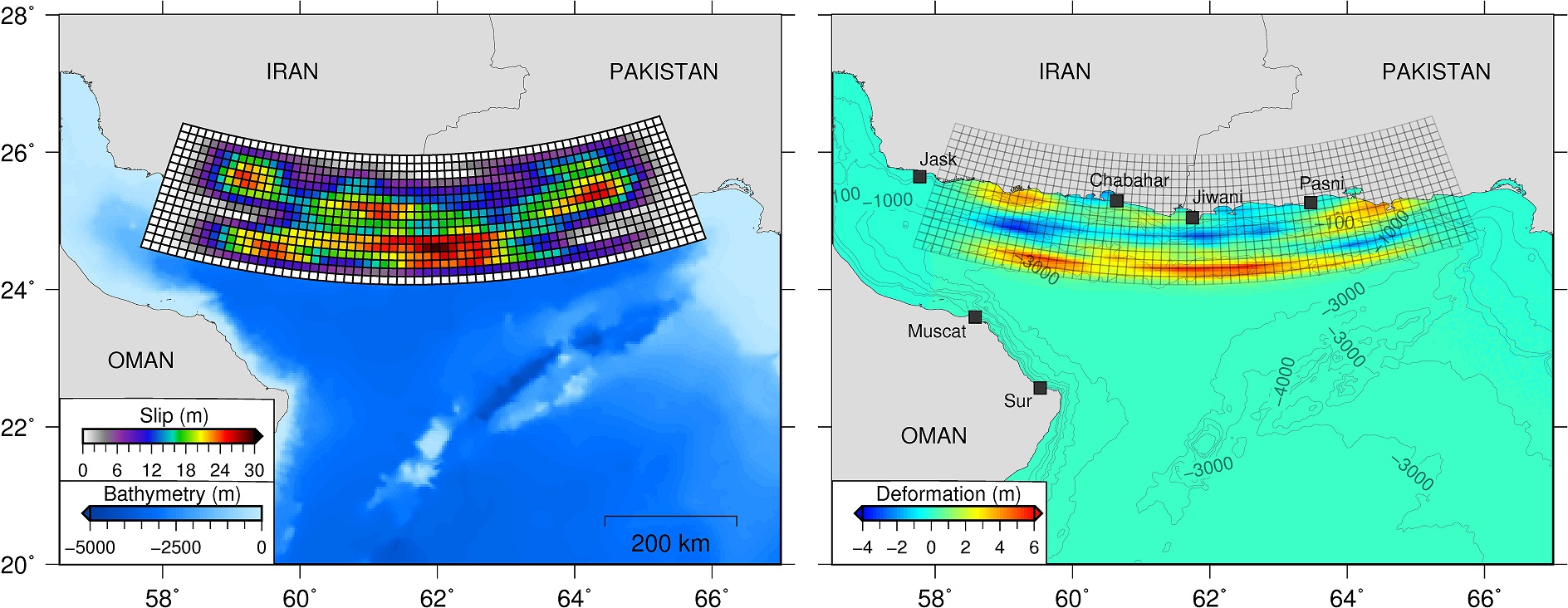}
\caption{\small\em An example of a $k^{-2}$ slip model for the \textsc{Makran} subduction zone (left panel) and the resulting vertical seafloor deformation (right panel). The non-planar mesh is the rupture geometry.}
\label{fig:slp}
\end{figure}

To define rupture scenarios, a non-planar fault geometry (Figure~\ref{fig:mak}) is constructed by modifying the plate interface and deformation front from \cite{GRL:GRL50374}. The southwest apex of the fault is located at $24.62^{\circ}$N and $57.68^{\circ}$E. The width of the fault is $210$ $\mathsf{km}$, which correspond to the limit of significant offshore seismicity \citep{GRL:GRL50374}. \cite{GRL:GRL50374} presented three potential rupture scenarios in the \textsc{Makran} region including the full length of the \textsc{Makran} subduction zone, eastern \textsc{Makran} segment and \textsc{Sistan} suture zone to \textsc{Little Murray} Ridge. They considered a coseismic slip of $10$ $\mathsf{m}$ for each scenario. Motivated by this work and considering a range of uncertainty, $100$ random heterogeneous $k^{\,-2}$ slip distributions are generated for hypothetical earthquakes using $10\ \pm\ 1$ $\mathsf{m}$ of mean coseismic slip and $K\ =\ 0.5\ \pm\ 0.2\,$. Figure~\ref{fig:slp} shows one of $k^{\,-2}$ slip distributions for the \textsc{Makran} subduction rupture, for instance.   
\subsection{Tsunami numerical modeling}

To provide the initial condition for tsunami modeling of earthquake scenarios, the seabed static deformation is computed using the \textsc{Okada} solution \citep{Okada85} (Figure~\ref{fig:slp}). We simulate tsunamis using the \textsc{Comcot} numerical model \citep{Liu1998} which uses explicit staggered leapfrog finite difference schemes to solve both linear and non-linear shallow water equations in both spherical and \textsc{Cartesian} coordinates. \textsc{Comcot} is a well-known and validated numerical tool that is used widely for investigating near-field and far-field tsunamis \citep[\emph{e.g.}][]{Wang2005, Wang2006, doi:10.1785/0120090211, doi:10.1093/gji/ggu297}. All simulations are performed for a total run time of $10$ $\mathsf{h}$ with a time discretization step of $2$ $\mathsf{s}$. The bottom friction is considered in the simulations. Nonlinear shallow water equations are discretized in spherical coordinates. Tsunami inundation on dry land is not computed due to the lack of high-resolution local bathymetry and topography data. Therefore, we have to use a global bathymetry grid for our purpose. The \textsc{Gebco} $30$ $\mathsf{arcsec}$ bathymetry data \citep{Smith1956, doi:10.1080/01490410903297766}, available at \url{http://www.gebco.net/}, is used for our simulations. 
\subsection{Tsunami hazard assessment}

\begin{figure}
\centering
\includegraphics[scale=0.32]{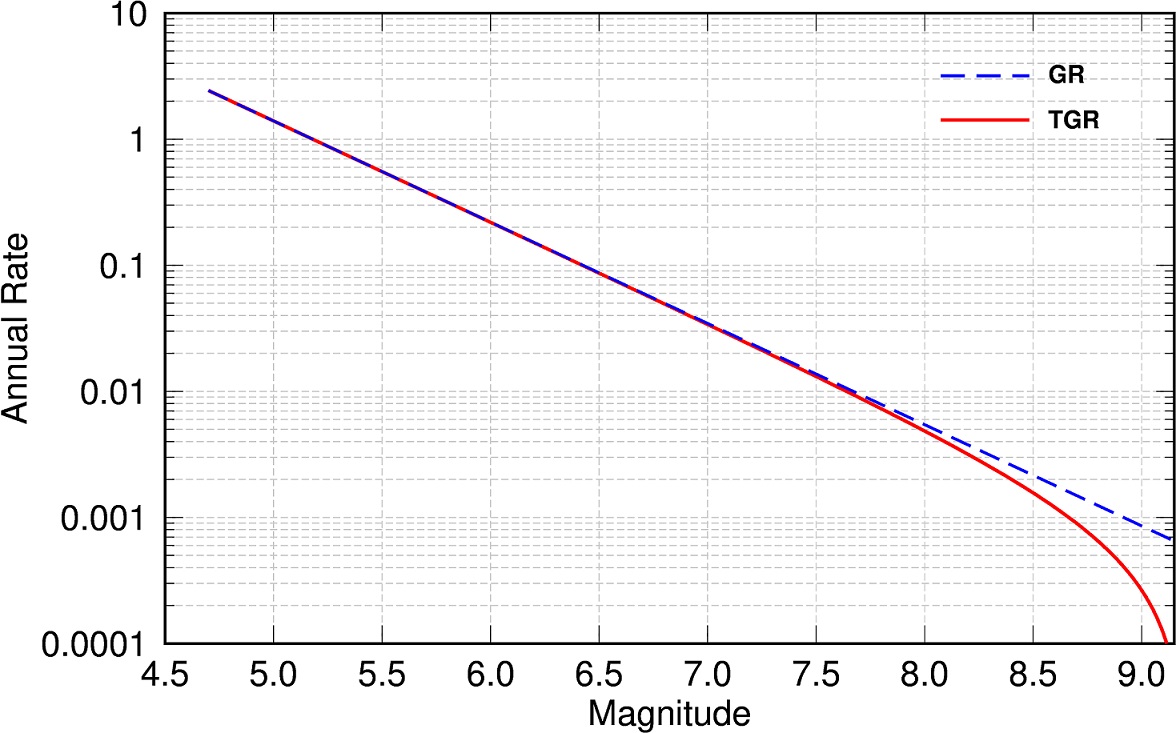}
\caption{\small\em Truncated \textsc{Gutenberg}-\textsc{Richter} (TGR, solid red line) and \textsc{Gutenberg}-\textsc{Richter} (GR, dashed blue line) relationships for the seismicity of \textsc{Makran} subduction zone between 1926-2016 from the ISC Catalog.}
\label{fig:gtr}
\end{figure}

The run-up distributions along the shorelines of \textsc{Iran}, \textsc{Pakistan} and \textsc{Oman} resulted from tsunami modeling of heterogeneous slip models are used to assess the probabilistic tsunami hazard. Computing annual rate of earthquakes as tsunami generators is required in PTHA. The truncated \textsc{Gutenberg}--\textsc{Richter} relation \citep{Cosentino1977:, Weichert1980} is applied to compute the annual number of the earthquakes (Figure~\ref{fig:gtr}) using events from 1926 -- 2016 driven from the ISC catalog (Figure~\ref{fig:mak}). Applying the maximum likelihood method \citep{Weichert1980}, the resulting magnitude of completeness is $4.7$ and the $b-$value $0.8$. The $b-$value is the slop of line in the \textsc{Gutenberg}--\textsc{Richter} relation and represents the relative abundance of large to small events. We consider a maximum magnitude of $9.2$ based on \cite{GRL:GRL50374}.

The occurrence of tsunamigenic earthquake scenarios is assumed to be a \textsc{Poissonian} process. The probability of tsunami wave amplitude $\zeta$ exceeding a specific value $\zeta_{\,c}$ at a given coastline in a time period $T$ for a total number $N$ of tsunamigenic sources can be wrriten as
\begin{equation} \label{eq:Pb}
  \mathbb{P}\,(\zeta\ \geq\ \zeta_{\,c})\ =\ 1\ -\ \prod_{i\,=\,1}^{N}\bigl[\,1\ -\ (1\ -\ \ue^{-\,v_{\,i}\,T})\,\mathbb{P}\,(\zeta\ \geq\ \zeta_{\,c}\;\vert\;S_{\,i})\,\bigr]\,.
\end{equation}
where $v_{\,i}$ is the annual occurrence rate of $i$\up{th} scenario $S_{\,i}$ and $\mathbb{P}\,(\zeta\ \geq\ \zeta_{\,c}\;\vert\,S_{\,i})$ is the probability that tsunami wave amplitude $\zeta$ resulted from $i$\up{th} scenario $S_{\,i}$ exceeds the given value $\zeta_{\,c}$ which is $1$ if $\zeta\ \geq\ \zeta_{\,c}$ and $0$ if $\zeta\ <\ \zeta_{\,c}\,$.
%
%
\section{Results}

\begin{figure}
\centering
\includegraphics[scale=2.7]{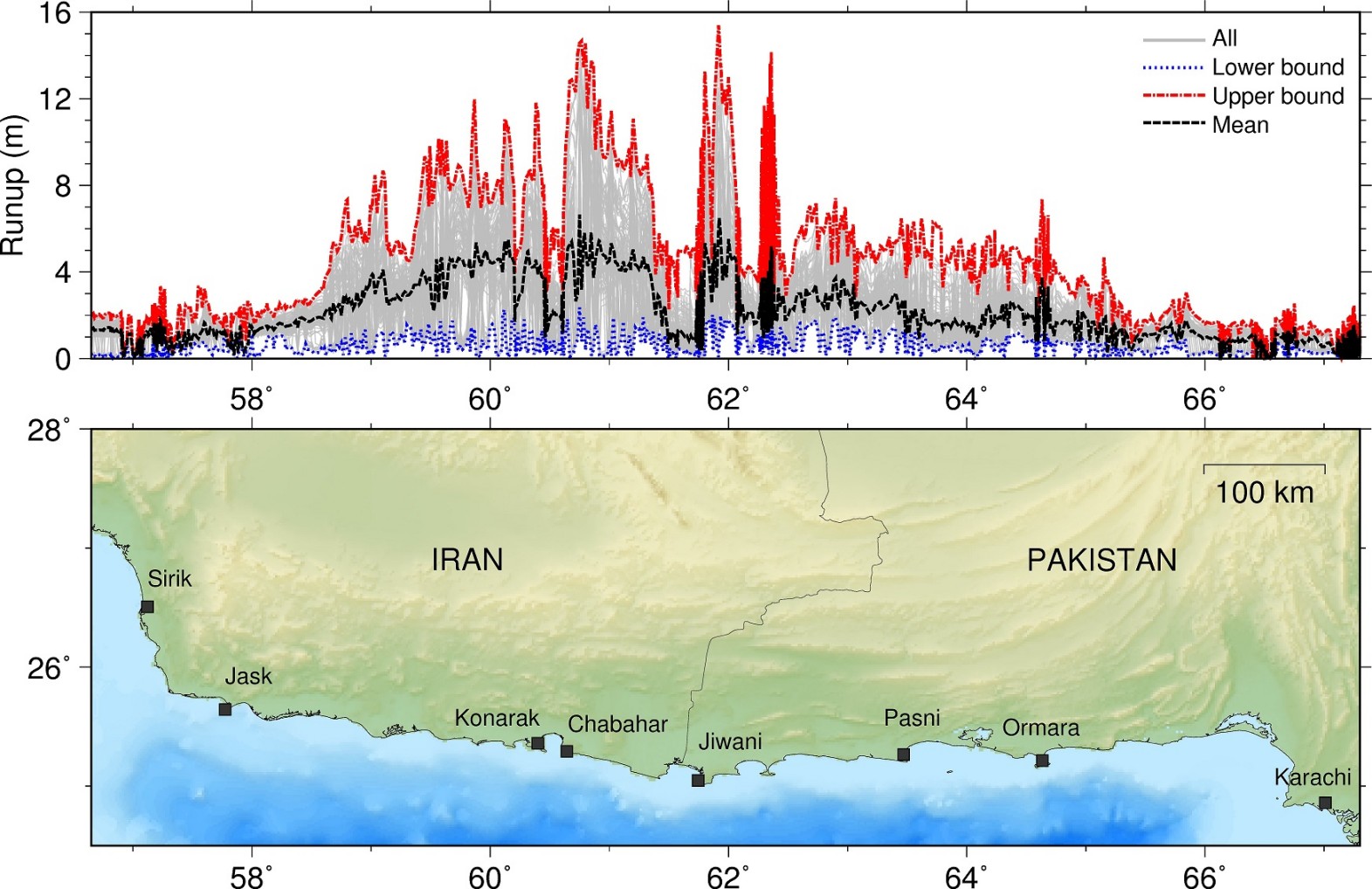}
\caption{\small\em Run-up distributions simulated in this study (gray lines) and mean (black dashed line), lower (blue dotted line) and upper (red dot-dashed line) run-up bounds along the coastlines of \textsc{Iran} and \textsc{Pakistan}.}
\label{fig:irpa}
\end{figure}

\begin{figure}
\centering
\includegraphics[scale=2.7]{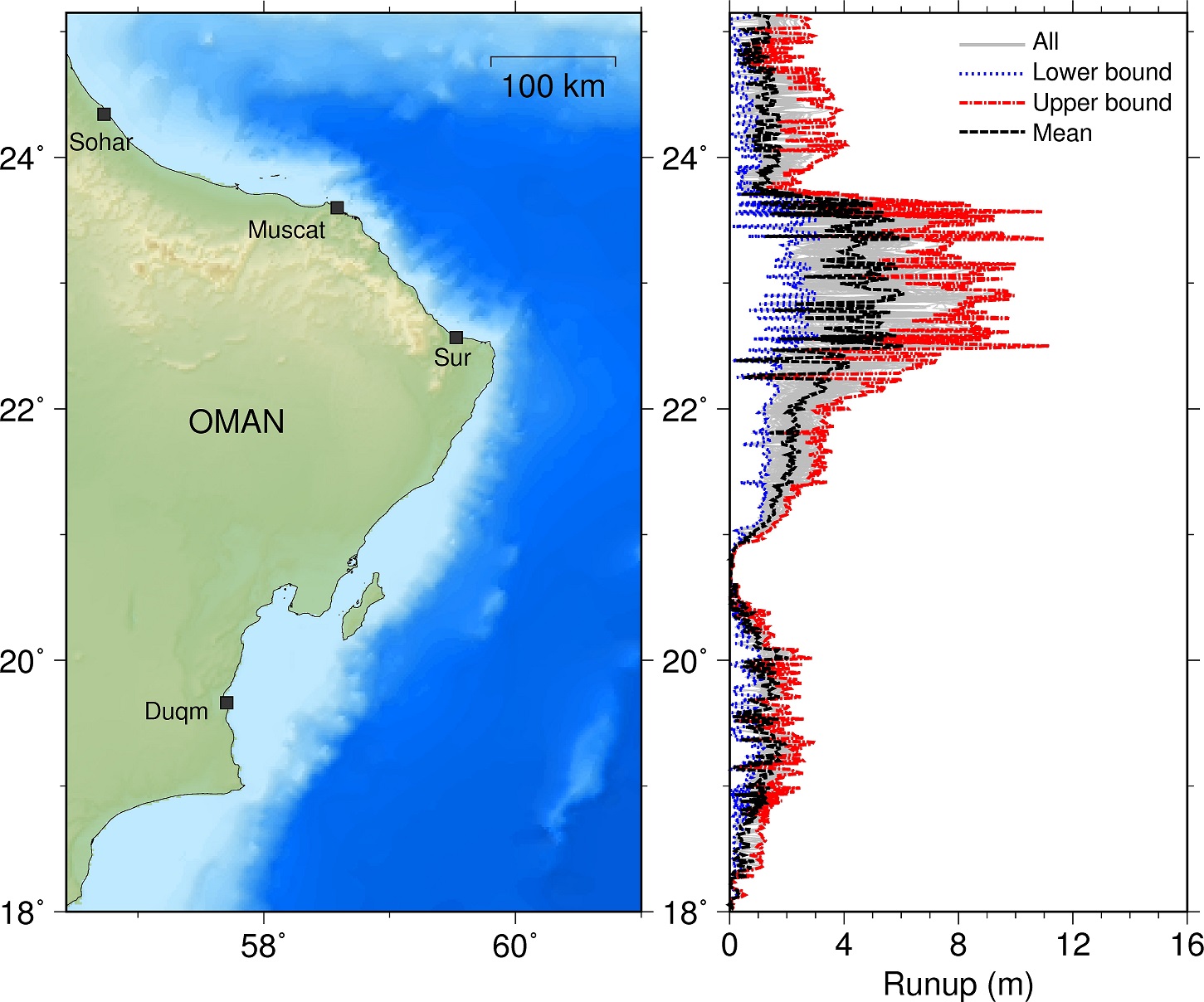}
\caption{\small\em Run-up distributions simulated in this study (gray lines) and mean (black dashed line), lower (blue dotted line) and upper (red dot-dashed line) run-up bounds along the coastline of \textsc{Oman}.}
\label{fig:om}
\end{figure}

Figure~\ref{fig:irpa} shows run-up distributions from all scenarios along the coastlines of \textsc{Iran} and \textsc{Pakistan}. The computed mean, lower and upper run-up bounds along the coastlines of \textsc{Iran} and \textsc{Pakistan} are also represented in Figure~\ref{fig:irpa}. The same results for the coastline of \textsc{Oman} are shown in Figure~\ref{fig:om}. The mean run-up along the shores of \textsc{Iran} and \textsc{Pakistan} ranges between $0$ and $7$ $\mathsf{m}$, while it changes between $0$ and $6$ $\mathsf{m}$ along the \textsc{Oman} coast. The maximum run-up reaches a height of $\sim 16$ $\mathsf{m}$ along the shores of \textsc{Iran} and \textsc{Pakistan} and $\sim 12$ $\mathsf{m}$ for the \textsc{Oman} shoreline. The run-up attenuates to the West and East of the \textsc{Iran} -- \textsc{Pakistan} shoreline. The highest run-up values along the shore are observed near the ports of \textsc{Chabahar} and \textsc{Jiwani}. In the case of \textsc{Oman} coast, an area between \textsc{Muscat} and South of \textsc{Sur} encounters the largest wave heights. Run-up heights attenuate to the South of \textsc{Oman} coastline, especially around the \textsc{Masirah Island}. Table~\ref{table:cities} shows the estimated mean and maximal run-up heights in some major coastal cities along the shores of \textsc{Iran}, \textsc{Pakistan} and \textsc{Oman}.

%
\begin{table}
\centering
\small
\renewcommand{\arraystretch}{1.45}
\begin{tabular}{l c c}
\hline
\multicolumn{1}{l}{\textit{City}} &
\multicolumn{1}{c}{\textit{Mean estimated run-up (m)} \quad\quad\quad\quad} &
\multicolumn{1}{c}{\textit{Maximal run-up (m)}} \\
\hline
\textsc{Jask} \quad\quad\quad\quad & $1$ \quad\quad\quad\quad & $2$ \\
\textsc{Konarak} \quad\quad\quad\quad & $4$ \quad\quad\quad\quad & $11$ \\
\textsc{Chabahar} \quad\quad\quad\quad & $5$ \quad\quad\quad\quad & $10$ \\
\textsc{Jiwani} \quad\quad\quad\quad & $2$ \quad\quad\quad\quad & $5$ \\
\textsc{Pasni} \quad\quad\quad\quad &  $3$ \quad\quad\quad\quad & $6$ \\
\textsc{Ormara} \quad\quad\quad\quad & $4$ \quad\quad\quad\quad & $8$ \\
\textsc{Muscat} \quad\quad\quad\quad & $5$ \quad\quad\quad\quad & $8$ \\
\textsc{Sur} \quad\quad\quad\quad & $5$ \quad\quad\quad\quad & $9$ \\
\hline
\end{tabular}
\bigskip
\normalsize
\caption{\small\em Estimated mean and maximal run-up heights in major coastal cities along the coastlines of \textsc{Iran}, \textsc{Pakistan} and \textsc{Oman}.}
\label{table:cities}
\end{table}

\begin{figure}
\centering
\includegraphics[scale=0.2]{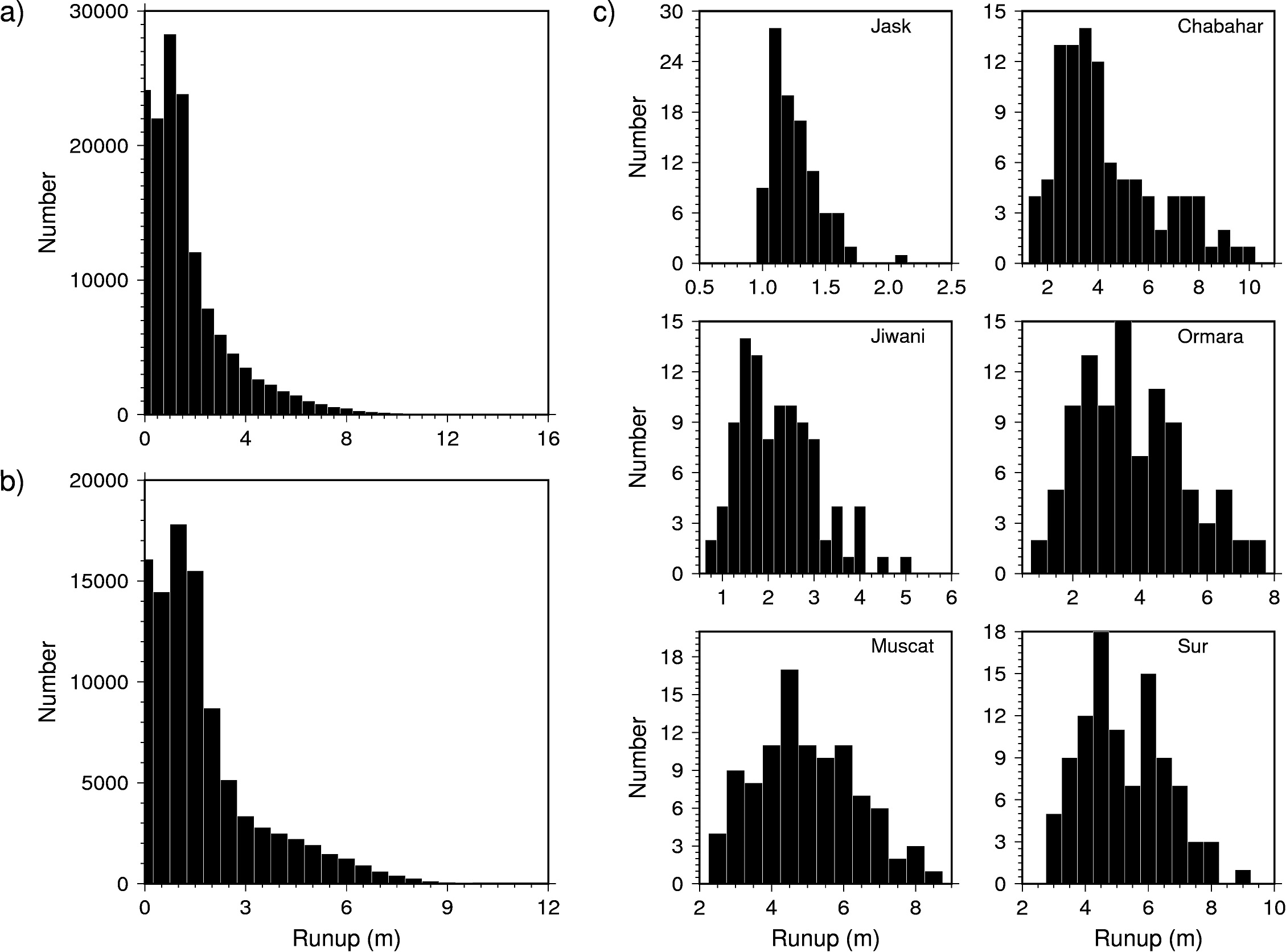}
\caption{\small\em Historgams of run-up heights along the coastlines of \textsc{Iran} -- \textsc{Pakistan} (a) and \textsc{Oman} (b) and for selected ports (c).}
\label{fig:hist}
\end{figure}

Figure~\ref{fig:hist} shows histograms of run-up heights along the coastlines of \textsc{Iran}, \textsc{Pakistan} and \textsc{Oman} and at several selected ports. About $90~\%$ of tsunami waves cause a run-up height up to $4$ $\mathsf{m}$. Among the selected locations, \textsc{Chabahar} is impacted by a larger number of higher run-up heights. \textsc{Jask} is exposed to the least hazard with maximum wave heights mainly from $1$ to $2$ $\mathsf{m}$.

\begin{figure}
\centering
\includegraphics[scale=0.285]{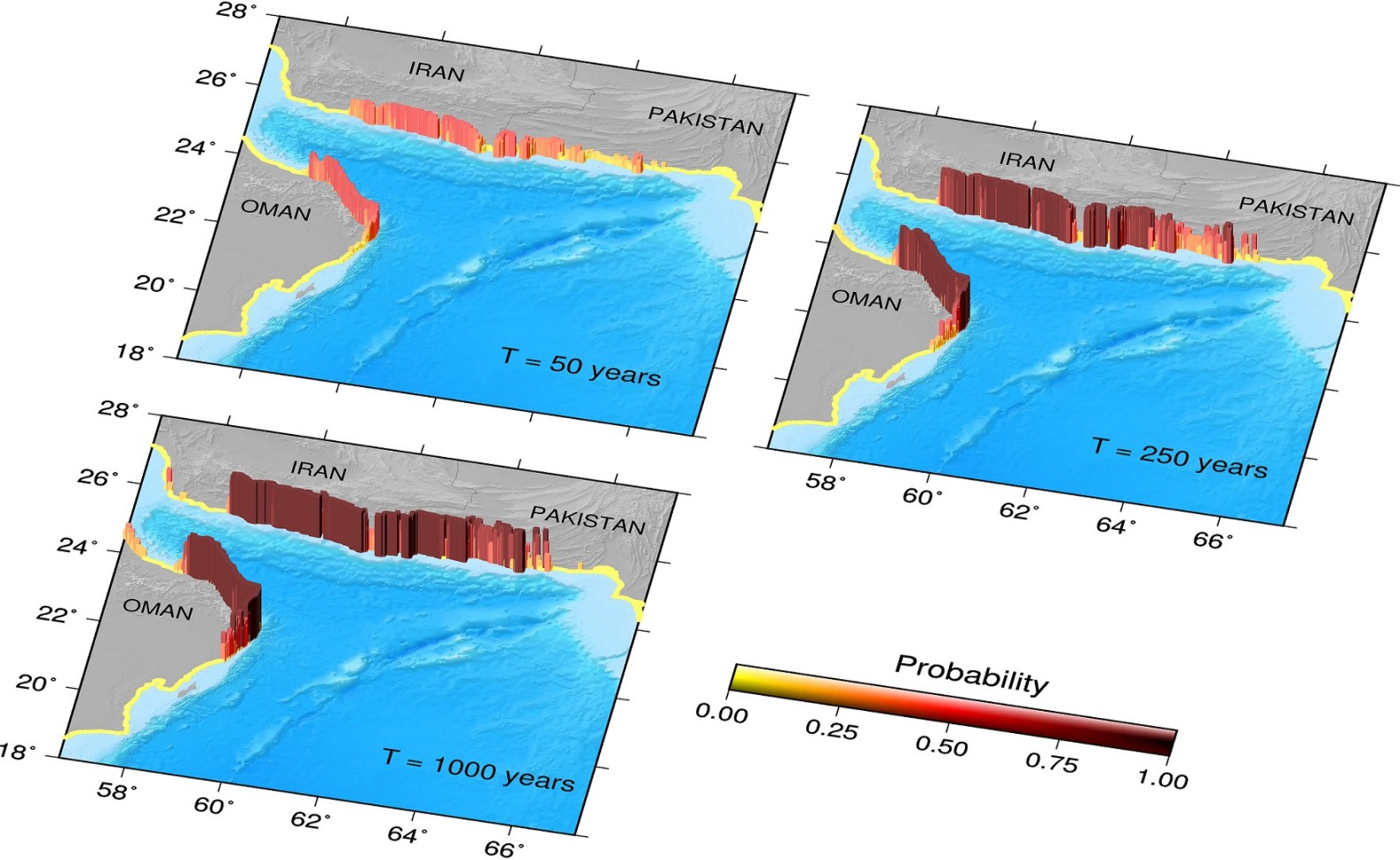}
\caption{\small\em Probability of exceending $3$ $\mathsf{m}$ in time periods of $50$, $250$ and $1\,000$ \textsf{years} along the coasts of \textsc{Iran}, \textsc{Pakistan} and \textsc{Oman}.}
\label{fig:prb_irpa}
\end{figure}

\begin{figure}
\centering
\includegraphics[scale=1.8]{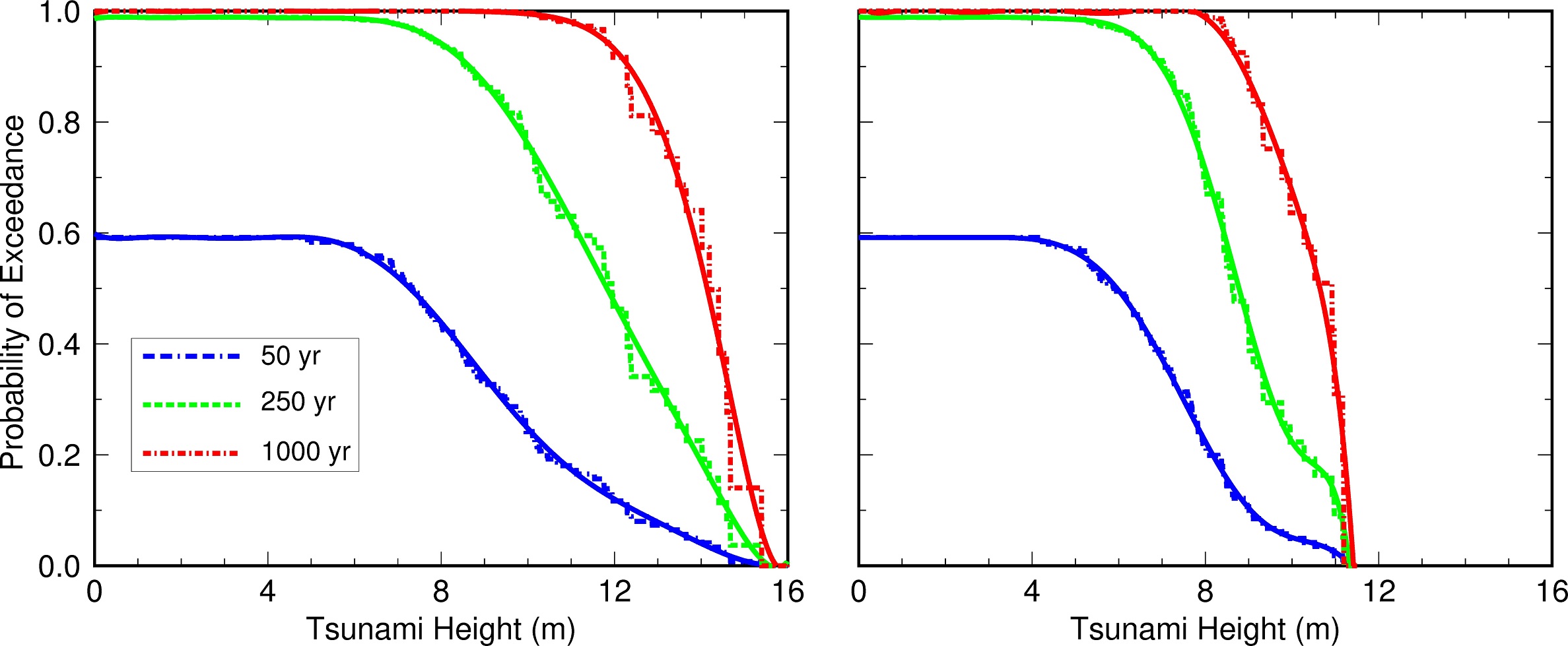}
\caption{\small\em Probability of exceending different wave heights at one or more locations along the coastlines of \textsc{Iran} -- \textsc{Pakistan} (left panel) and \textsc{Oman} (right panel). Solid lines indicate robust fitting.}
\label{fig:pbr_om}
\end{figure}

\begin{figure}
\centering
\includegraphics[scale=1.8]{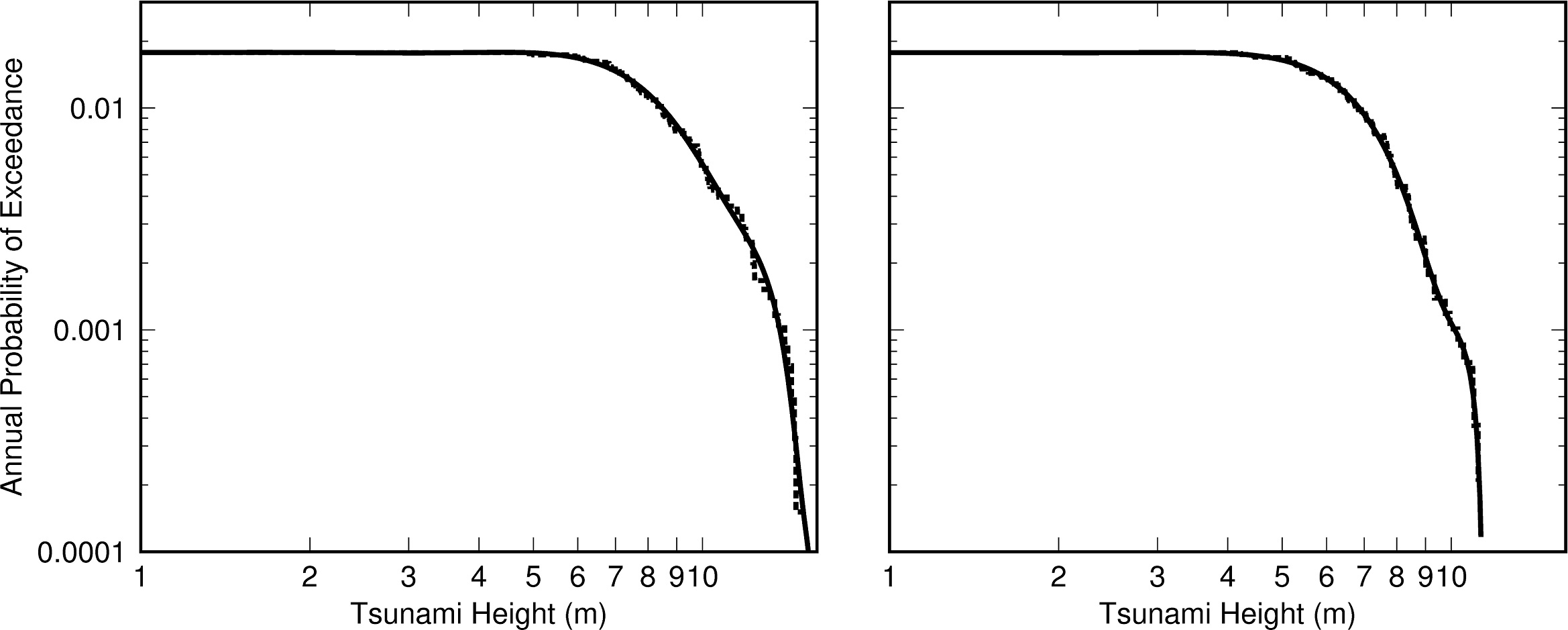}
\caption{\small\em Annual probability of exceending different wave heights at one or more locations along the coastlines of \textsc{Iran} -- \textsc{Pakistan} (left panel) and \textsc{Oman} (right panel). Solid lines indicate robust fitting.}
\label{fig:an_pbr}
\end{figure}

The results of PTHA are presented in Figures~\ref{fig:prb_irpa} -- \ref{fig:an_pbr}. Figure~\ref{fig:prb_irpa} shows the probability of tsunami wave amplitude exceeding $3$ $\mathsf{m}$ evaluated for coastlines of \textsc{Iran}, \textsc{Pakistan} and \textsc{Oman} in time periods of $50$, $250$ and $1\,000$ \textsf{years}. Distributions of the exceedance probability in $50$, $250$ and $1\,000$ \textsf{years} illustrate a relatively similar pattern along the shorelines. The probability of exceeding $3$ $\mathsf{m}$ increases with time. However, its changes in the West and East of \textsc{Iran} -- \textsc{Pakistan} shoreline and West of \textsc{Muscat} are insignificant. For a region between \textsc{Jask} and \textsc{Ormara} along the \textsc{Iran} -- \textsc{Pakistan} shoreline, the probability of exceedance is high. It ranges from $0$ to $0.56$ in $50$ \textsf{years} and reaches to $1$ in $1\,000$ \textsf{years} for most locations between \textsc{Jask} and \textsc{Ormara}. The exceedance probability is significant between \textsc{Muscat} and \textsc{Sur}, ranging from $0$ to $0.99$ ($\sim$ 1) in $250$ \textsf{years}. It decreases to less than $0.11$ to the West of \textsc{Muscat}.

Figure~\ref{fig:pbr_om} shows the probability of exceeding different wave heights at one or more locations along the coastlines of \textsc{Iran} and \textsc{Pakistan} and \textsc{Oman} in time periods of $50$, $250$ and $1\,000$ \textsf{years}. The same plots for the annual probability are displayed in Figure~\ref{fig:an_pbr}. For both coastlines of \textsc{Iran} -- \textsc{Pakistan} and \textsc{Oman}, the value of exceedance probability for $3$ $\mathsf{m}$ is about $0.6$ in $50$ \textsf{years} and $1$ in $250$ and $1\,000$ \textsf{years}. The probability that tsunami wave height exceeds $9$ $\mathsf{m}$ at at least one location along the \textsc{Iran} -- \textsc{Pakistan} shoreline is $0.33$, $0.86$ and $1$ during $50$, $250$ and $1\,000$ \textsf{years}, respectively. The same values for the coast of \textsc{Oman} are $0.11$, $0.43$ and $0.9$. The annual exceedance probability for tsunami wave heights between $1$ and $5$ $\mathsf{m}$ is a constant value of $0.018$. It falls exponentially for higher tsunami heights.
%
%
\section{Discussion and Conclusions}

Taking into account a range of near-field source models, our tsunami simulations show that maximum run-up reaches to a height of $16$ $\mathsf{m}$ along the coastlines of \textsc{Iran} and \textsc{Pakistan}. The minimum bound of run-up exhibits a relative uniform distribution which can match with run-up from a uniform slip distribution \citep{Ruiz2015}. The results of our simulations show that near-field run-up along the coastlines depends on the slip distribution of tsunamigenic events. Different slip models result in different patterns of run-up along the shores, especially for the zones located parallel to the rupture area. The patterns of run-up from non-uniform slip distributions are more complex than uniform slip distributions. Moreover, the resolution of bathymetry can largely affect run-up heights. Clearly, the higher resolution bathymetry grid has, the more accurate run-up computation is. Besides the different slip distributions, the non-uniform shape of the \textsc{Iran} -- \textsc{Pakistan} coastline due to the presence of different coastal features influences the pattern of run-up. A high variability of run-up heights can be seen between $58.5^{\circ}$E and $65.3^{\circ}$E, which corresponds to the concentration of slip over the rupture area. This part of the shore is the most affected area by tsunamis. To the West and East of \textsc{Iran} -- \textsc{Pakistan} shoreline, the dependency of run-up to the heterogeneity of slip relatively decreases. It reflects a lower uncertainity of run-up in these areas.

For the coast of \textsc{Oman}, the variability of run-up is high in an area between \textsc{Muscat} and \textsc{Sur}. The uncertainty of run-up showed by lower and upper run-up bounds is lower in other parts of the \textsc{Oman} shore. To the South of \textsc{Oman}'s eastern tip, run-up heights are highly attenuated around the \textsc{Masirah Island} because of the tsunami waves scattering. However, it seems that run-up values amplify locally to the South of the \textsc{Masirah Island} due to multiple reflections of tsunami waves.

The probabilistic tsunami hazard for the central part of \textsc{Iran} -- \textsc{Pakistan} shoreline and the area between \textsc{Muscat} and \textsc{Sur} is quite considerable. Tsunami waves deplete their energy mainly along those areas. The location of western and eastern parts of \textsc{Iran} -- \textsc{Pakistan} shoreline and the West of \textsc{Muscat} with respect to the extent of slip on the fault area causes very weaker run-up heights and thus lower levels of tsunami hazard in comparison to the central areas. Scattering of tsunami waves around \textsc{Oman}'s eastern tip can dissipate tsunami waves energy and cause attenuated run-up to the South of \textsc{Oman}. While the tsunami hazard probability increases with time along the most affected areas, it stays nearly constant in other areas. The short-term ($50$ \textsf{years}), mid-term ($250$ \textsf{years}) and long-term ($\sim 1\,000$ \textsf{years}) tsunami hazard along the main affected areas are significant which make them highly vulnerable to tsunami hazard. Other areas are less susceptible to tsunami hazard. Tsunami hazard variability with time depends on size and magnitude of tsunamigenic scenarii. This study presumes a certain size of the \textsc{Makran} rupture area with a limited range of magnitudes. Considering different sizes of rupture areas with different magnitudes will change the results. In general, \textsc{Iran} is more vulnerable than \textsc{Pakistan} and \textsc{Oman} to tsunami hazard from the entire \textsc{Makran} subduction zone. High run-up heights are distributed along a broader area for the \textsc{Iranian} part of the shoreline.
 
This study takes into account the effect of slip heterogeneity on tsunami hazard along the shorelines of \textsc{Iran}, \textsc{Pakistan} and \textsc{Oman} considering a range of uncertainty for mean slip value and slip distribution. Future works should consider the effects of other factors on run-up along those coastlines (\emph{e.g.} fault parameters, the dynamic bottom motion, other near-field and far-field tsunamigenic sources, \emph{etc.}). Besides, high resolution site-specific bathymetric/topographic maps are required to improve the results of our study and to compute the tsunami inundation of dry land. 

The high level long-term tsunami hazard of \textsc{Makran} subduction zone indicates the need of preparedness for future events especially by developing a regional tsunami early warning system for \textsc{Makran} coastlines. Special attention needs to be paid to the central part of \textsc{Iran} -- \textsc{Pakistan} shoreline and the area between \textsc{Muscat} and \textsc{Sur}. Local authorities might be interested to use the results of this study alongside with other relevant publications in order to increase the protection level of coastal facilities. The \textsc{Makran} subduction zone is a common threat for different countries, especially \textsc{Iran}, \textsc{Oman}, \textsc{Pakistan} and \textsc{India}. Collaborative studies, sharing data and tsunami awareness education can constructively  mitigate the \textsc{Makran} tsunami hazard.
%
%
%
\bigskip
\renewcommand{\bibsection}{\section*{References}}
\bibliography{Bibl}
\bigskip
%
%

\end{document}